# Pole-Zero Identification: Unveiling the Critical Dynamics of Microwave Circuits Beyond Stability Analysis

J.M. Collantes[1], L. Mori[1], A. Anakabe[1], N. Otegi[1], I. Lizarraga[1], N. Ayllón[2], F. Ramírez[3], V. Armengaud[4], G. Soubercaze-Pun[4]

1 – Electricity and Electronics Dpt. University of the Basque Country (UPV/EHU), Bilbao, Spain
2 – European Space Agency, Noordwijk, The Netherlands
3 – Dpt. of Communications Engineering, University of Cantabria, Santander, Spain
4 – French Space Agency (CNES), Toulouse Space Centre, Toulouse, France

Pole-zero identification refers to the obtaining of the poles and zeros of a linear (or linearized) system described by its frequency response. This is usually done using optimization techniques (such as least squares, maximum likelihood estimation, or vector fitting) that fit a given frequency response of the linear system to a transfer function defined as the ratio of two polynomials [1]-[2]. This kind of linear system identification in the frequency domain has numerous applications in a wide variety of engineering fields (such as mechanical systems, power systems and Electromagnetic Compatibility). In the microwave domain, rational approximation is increasingly used to obtain black-box models of complex passive structures for model order reduction and efficient transient simulation. An extensive bibliography on the matter can be found in [3]-[6]. In this paper we will focus on a different application of pole-zero identification. We will review the different ways in which pole-zero identification can be applied to nonlinear circuit design (for power amplifier stability analysis and beyond). We will give a comprehensive view on recent approaches through illustrative application examples. Other uses of rational approximation techniques are beyond the scope of this paper.

In the context of microwave amplifier design, pole-zero identification was introduced in 2001 as an alternative method to analyze the stability of an active circuit under small and large-signal excitations [7]. It provided a simple and intuitive way to solve the generalized eigenvalue problem without the need to access the Jacobian of the system [8] (not available in commercial simulators) or to the internal nodes of the non-linear models of active devices [9] (very often not available for IP protection). Thus, it is perfectly fitted for use in combination with conventional microwave simulators as Advanced Design System (ADS) or Microwave Office (MWO). Pole-zero identification has been increasingly used ever since to analyze the stability of active microwave circuits, mainly amplifiers [10]-[16], but also oscillators [17]-[18], frequency dividers [19]-[20] or even non-Foster circuits [21]. Perhaps the most attractive features of pole-zero identification for stability are its simplicity (it is actually a probe-based method) and the graphical and intuitive nature of the results: whenever a pair of complex conjugate poles is found lying on the Right-Half Plane (RHP), i.e. poles with positive real part, an oscillation will begin to grow from the steady state. The basic steps for the stability analysis are equivalent for both small-signal and large-signal regimes. Obviously, the simulation engines differ (simple AC analysis in one case and harmonic balance with conversion matrix in the other) but the process of probing the circuit, fitting a frequency response and monitoring the resulting poles in a pole-zero map are completely analogous.

Getting the pole-zero map of a linearized system provides more information than just a pass/fail stability analysis. As a matter of fact, pole-zero identification in the microwave active circuit context has evolved to become a useful tool to obtain relevant information regarding the non-linear dynamics of any kind of active device. In this paper we will provide several examples on how pole-zero identification is being used to increase stability margins in power amplifiers, to obtain the operation bands in autonomous circuits or to get the optimum stabilization solution in multi-transistor circuits. We will also review recent developments in the application of pole-zero identification techniques that can foster their use in active circuit analysis and design.

**Fundamentals of stability analysis based on pole-zero identification**

Stability analysis based on pole-zero identification is essentially a probe-based technique. In its most basic version, the analysis is carried out in two main steps: A first one to simulate a frequency response in the microwave commercial simulator with the aid of a small-signal current or voltage source (the probe); and a second one to fit that frequency response to a transfer function, expressed as the ratio of two polynomials, using a pole-zero identification tool.

*Step 1: Obtaining the closed-loop frequency response*

The first step is obtaining a Single-Input Single-Output (SISO) closed-loop frequency response of the circuit linearized about its steady state regime. The steady state can be either a DC bias point or a large-signal periodic regime forced by the input drive. In order to get the closed-loop frequency response, the circuit is "probed" either by a small-signal current source connected at a circuit node or by a small-signal voltage source inserted in series into a circuit branch. The frequency of the source is swept along the band of analysis. When a current source is used, the closed-loop frequency response is given by the impedance seen by the probe at the connection node (Fig. 1a). In turn, when using a voltage source, the closed-loop frequency response is the total admittance presented to the probe (Fig. 1b). Note that other responses are also valid, such as trans-impedances, trans-admittances or voltage and current transfers as long as they represent a closed-loop response of the linearized system. The system being linear (or linearized), all closed-loop responses will provide the same set of poles except for exact pole-zero cancellations. This means that, in theory, we can probe the circuit at any node/branch and obtain the same stability information. We know that this is not completely true in practice. Actually, there can be nodes and branches in a circuit that are electrically isolated from part of the circuit's dynamics. Probing the circuit at those nodes or branches with very low sensitivity (lack of controllability and observability[1]) can lead to a pole-zero cancellation and an unstable pole might remain undetected. This can happen when probing at very low impedance nodes with the current source or at very low admittance branches with the voltage source [22]. A typical example is the case of odd-mode oscillations in power combining amplifiers. Probing with a current source at a power combination node leads to an exact pole-zero cancellation because the combination node is actually a virtual ground for the odd-mode signal. That is why, in multi-stage circuits, it is recommended performing the analysis on at least one node or branch per stage and preferably close to a transistor. We will come back to this point later in the text.

---

[1] In control theory, observability refers to the capability to infer internal states of a system from external outputs, while controllability is related to the ability to modify the state of a system acting from an external output. Both concepts are dual aspects of the same problem.

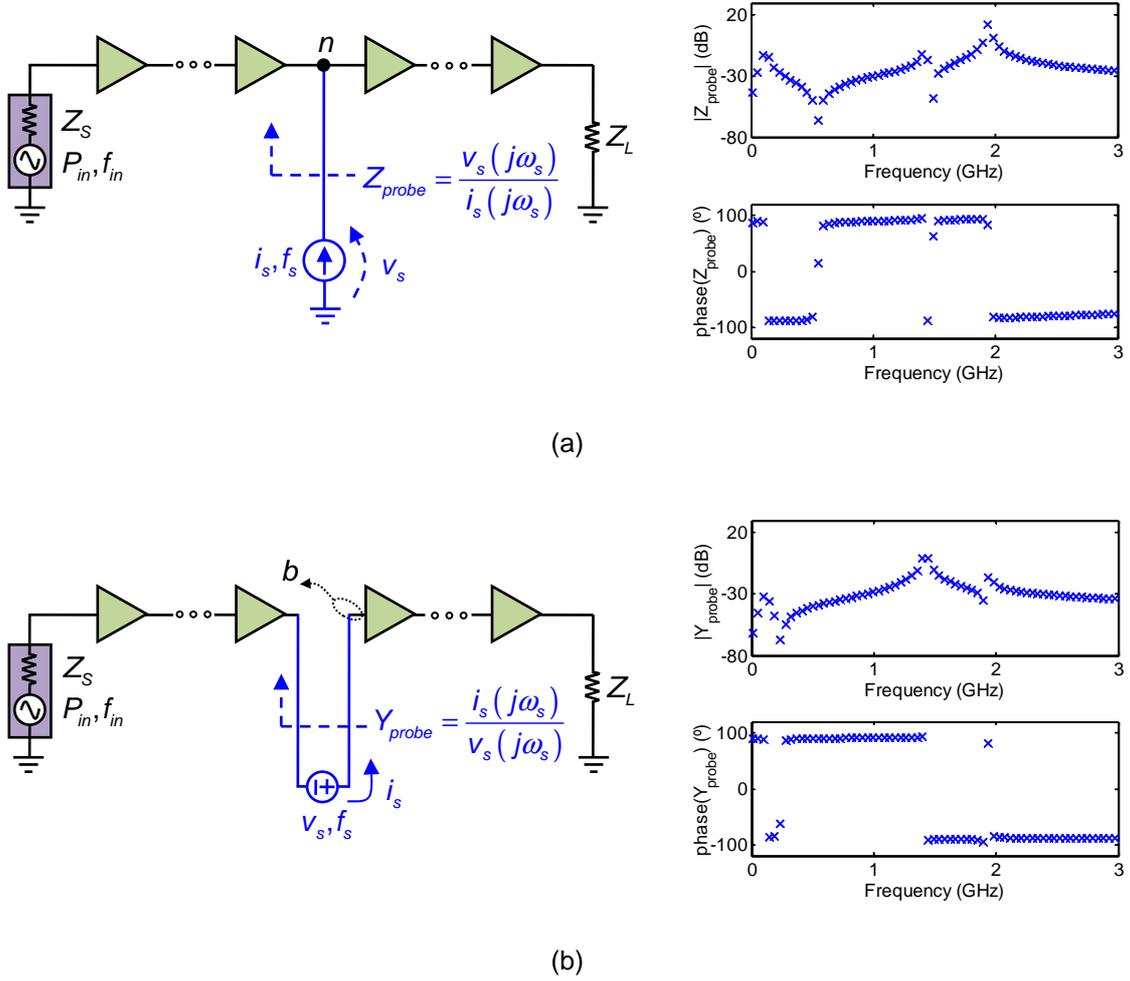

Fig. 1.  STEP 1: Obtaining a closed-loop frequency response for stability analysis (a) with a current source connected at node *n*: the closed-loop frequency response is given by the impedance seen by the probe at connection node *n*. (b) with a voltage source connected in series into branch *b*: the closed-loop frequency response is given by the total admittance presented to the probe.

If the steady state corresponds to a DC bias point or to a small-signal regime, the frequency sweep is carried out by a common AC analysis in the simulator. System poles are the eigenvalues of the linearized system and the frequency sweep should cover the entire band in which the active devices have gain (up to $f_{max}$). When the steady state is a periodic regime driven by a periodic large-signal input at a fundamental frequency $f_{in}$, a mixer-like analysis based on the conversion matrix algorithm has to be used to simulate the frequency response (this is called harmonic balance simulation with *small-signal mode* in ADS and *large-signal small-signal analysis* in MWO). In this case, the system poles correspond to the Floquet exponents that appear periodically with the fundamental frequency $f_{in}$ and, consequently, the frequency sweep of the analysis can be constrained to [0 - $f_{in}$/2] due to this periodicity.

*Step 2 – Obtaining the transfer function*

Once the closed-loop frequency response has been simulated in the commercial CAD tool, we can move on to the second step of the process. This consists in fitting the frequency response to a transfer function formulated as the ratio of two polynomials (Fig. 2). The fitting mechanisms implicitly approximate the delays introduced by the transmission lines to the polynomial representation. This polynomial approximation of the transmission lines has been demonstrated to work satisfactorily for amplifier designs up to Ka band and beyond [16]. Once the transfer function is obtained, we can plot the resulting poles and zeros on the complex plane to get graphical information on the system stability. In this context, model stability is not enforced in the identification process because we search for possible unstable poles. In the case shown in Fig. 2, a pair of complex conjugate poles lies on the RHP indicating that the system is unstable and an oscillation will begin to build up. Initially, this oscillation starts to grow at the frequency of the poles (≈1.4 GHz). Note that no quantitative values can be given for the final amplitude or frequency of the oscillation from a local stability analysis (in which the system has been linearized about a steady state).

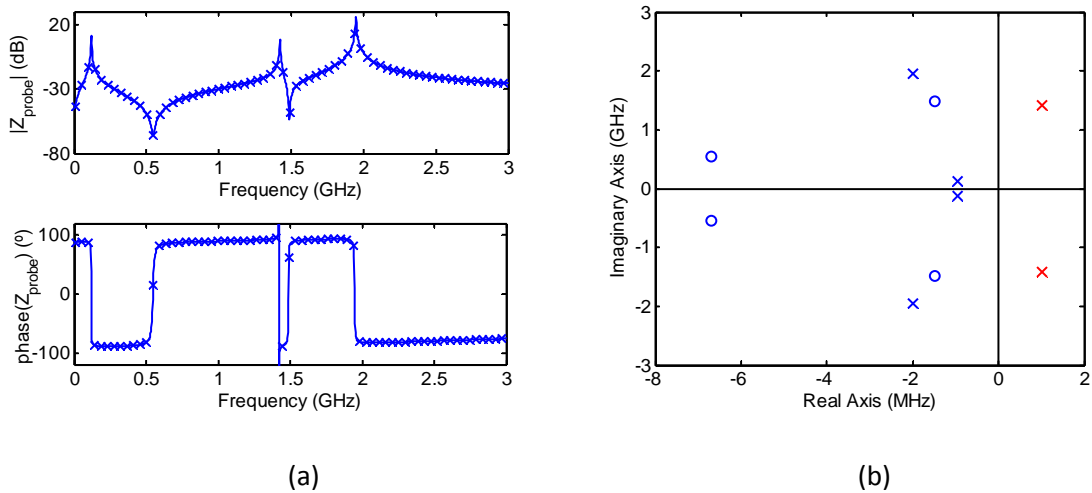

(a)                                                                (b)

Fig. 2. STEP 2: Fitting the closed-loop frequency response obtained in step 1 to a transfer function formulated as the ratio of two polynomials. Identification results: (a) Identified transfer function (solid line) superimposed to the original frequency response (crosses). (b) Associated pole-zero map. x: poles, o: zeros. Unstable complex conjugate poles are obtained at around 1.4 GHz indicating that the analyzed steady state is unstable and an oscillation starts to build up at 1.4 GHz.

As in any general linear identification process, in stability analysis the quality assessment of the identification is also a critical step because the order of the transfer function is *a priori* unknown. The goal is to obtain an approximation that prevents both under-modeling (a critical pole may be missing in the transfer function) and over-modeling (the order of the transfer function is unnecessarily high for the system dynamics). Consequences of under-modeling and over-modeling, in the context of stability analysis, are illustrated in Fig. 3 where a frequency response with a small unstable resonance is identified with different transfer function orders.

With an order n = 4 for both numerator and denominator, the instability is correctly detected by a pair of RHP complex conjugate poles that are quasi-cancelled by nearby zeros (Fig. 3a). This quasi-cancellation is indicating a low sensitivity to this dynamics from the observation port at which the probe has been connected, as stated before. The instability is missed if we use an order as low as n = 2 (Fig. 3b). With an order unnecessarily high, like n = 6, we run into an over-modeling problem with the presence of an additional pole-zero quasi-cancellation on the RHP (Fig. 3c). This kind of over-modeling quasi-cancellation has a numerical origin and can be located anywhere on the complex plane, not necessarily on the RHP. When they appear on the left-half plane (LHP) they are not problematic because they do not modify the stability results. In that case, they are normally ignored because the goal here is not to generate a numerically stable model for transient simulation[2], but to guarantee a reliable detection of critical poles (poles that cross to the RHP as some circuit parameter varies). However, over-modeling quasi-cancellations lying on the RHP are a major concern because they can be mistaken as physical poles obtained at a low-sensitivity node or branch, leading to wrong conclusions in terms of stability. A prime objective of the identification process is to discriminate critical poles that truly take part in the circuit dynamics from non-physical poles that may appear numerically due to over-modeling. This can be particularly challenging when analyzing a system with very rich dynamics in a wide frequency band. An obvious approach to tackle this issue is to perform the analysis at different nodes/branches until the RHP physical poles (if any) are detected with high sensitivity. This entails a clear unstable resonance, as the one in Fig. 2, with a pair of complex conjugate poles not quasi-cancelled by nearby zeros, which cannot be confused with numerical over-modeling. Another typical solution that does not require the analysis at other nodes/branches is based on reducing the complexity of the problem by cutting a broadband frequency response in narrow sub-bands. Actually, over-modeling numerical quasi-cancellations are very dependent on the bandwidth of the analyzed frequency range, while physical poles are consistently found at the same location whatever this bandwidth is. Therefore, a series of re-identifications in narrow sub-bands centered at the frequency of the critical quasi-cancellation are commonly carried out to verify/discard by visual inspection the physical origin of RHP quasi-cancellations. In [23] this concept was used to provide an automatic procedure to eliminate RHP over-modeling quasi-cancellations. Obviously, this kind of strategy is not intended for the extraction of a complete model, but only for the reliable detection of critical poles.

---

[2] Contrarily to the general rational approximation problem, in the context of stability analysis, the identification process does not need to produce a numerically stable rational model for a later simulation in a time domain simulator. In this sense, pole-zero identification for stability analysis is less demanding than black-box modeling of passive structures through rational approximations, which have to generate numerically stable models in large bands and guarantee stability and passivity.

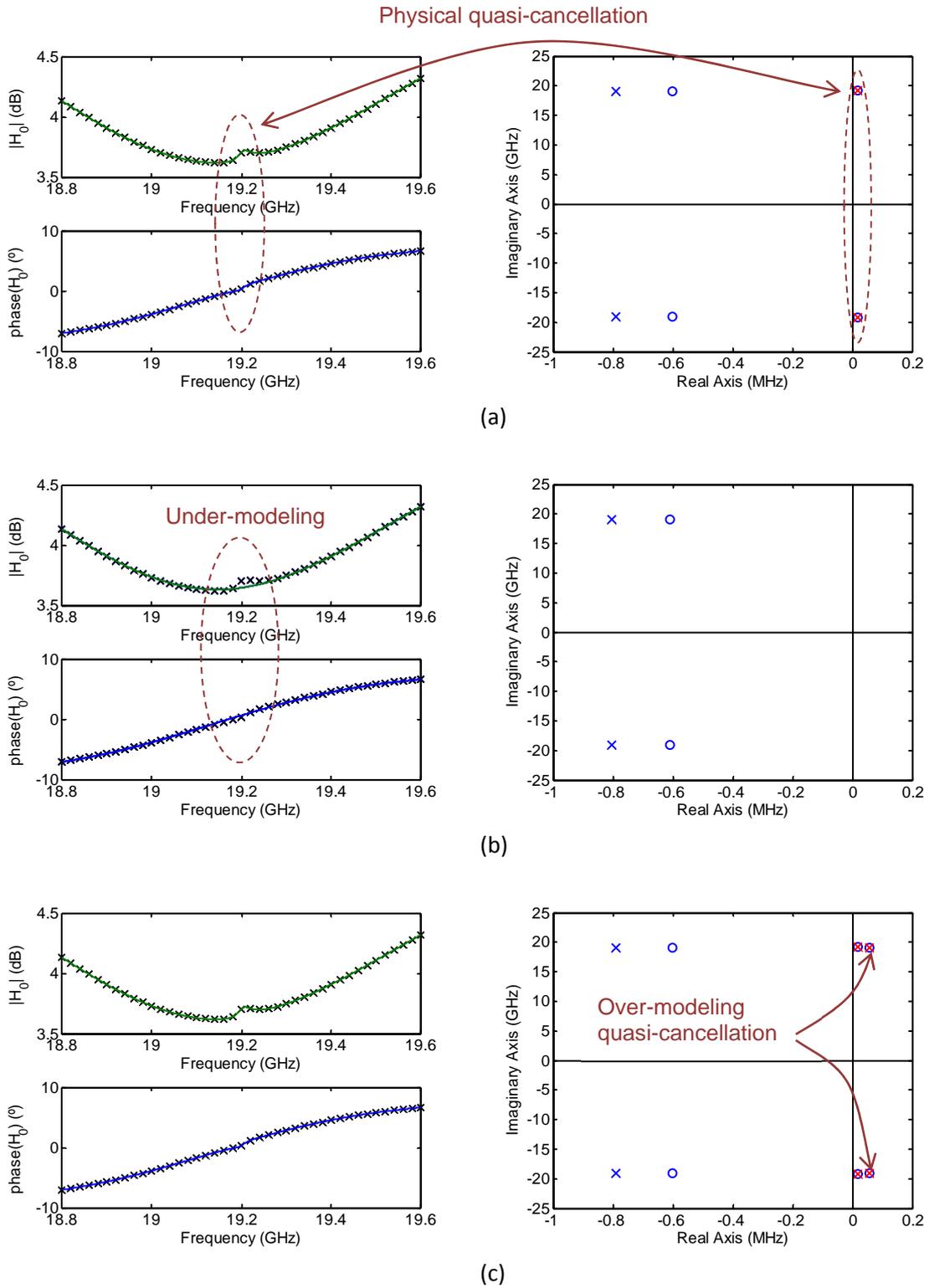

Fig. 3.  Identification results of a frequency response $H_0$ with different transfer function orders [23]. (a) n = 4, (b) n = 2 and (c) n = 6. Left graphs represent magnitude and phase of the frequency responses (original data with crosses and identified results with solid line). Right graphs are the corresponding pole-zero maps (×: poles, ○: zeros). Physical unstable poles are correctly detected with n = 4. The instability is missed with n = 2 and an over-modeling quasi-cancellation appears with n = 6.

Tools for linear system identification in the frequency domain are available as routines or toolboxes in numerical computing environments, such as *tfest.m* [24], *vectfit.m* [25] or *FDIDENT* [26] in Matlab or *frep2tf.sci* in Scilab [27]. There exist, as well, commercial tools specifically adapted for the pole-zero identification in the context of microwave amplifier stability, such as the *STAN Tool* [28]. Although less frequently employed in nonlinear circuit design, other performing and commercially available tools for rational model generation of passive multi-port networks such as *IdEM* [29] or *Sigrity Broadband SPICE* [30] could also be used (provided that the passivity condition is not enforced).

Finally, the reliability of the stability analysis depends more on the accuracy of the simulated frequency response than on the pole-zero identification process that follows. Lack of precise electrical models, incomplete circuit descriptions or numerical errors in the simulated frequency response (bad convergence, large numerical noise, truncation…) can lead to inaccurate or even non-physical frequency responses. The identification of invalid frequency responses can produce false pole-zero plots with wrong conclusions about the stability of the system.

**More than a pass/fail stability test**

Pole-zero identification is more than a simple pass/fail stability check. Obtaining the position of the poles and zeros on the complex plane and monitoring their evolution as some relevant circuit parameters are varied provide useful information that can be used for bifurcation analysis, transient control, optimum circuit stabilization and stability margin evaluation. We review here some examples in different contexts:

*Circuit stabilization*

A major value of pole-zero identification lies in its capability to guide designers to an optimal stabilization of the circuit. Pole-zero plots obtained at different nodes and pole trajectories on the complex plane are useful instruments that can help designers select the appropriate stabilization elements that ensure stable behavior with a minimum impact on the amplifier performance. A number of examples can be found in the literature. In [31], a control design in the frequency domain is applied to the stabilization of a varactor-based circuit. In [32], a systematic method was proposed to determine suitable topology, location and value of the stabilization elements in a multi-device microwave amplifier. Their methodology is based on detecting the sensitive nodes and/or branches of the circuit at which the inclusion of a stabilization network (shunt or series resistance, capacitance, inductance or a combination) is able to eliminate an undesired oscillation. This is done by analyzing the position of the zeros relative to the poles, obtained at different observation nodes or branches. If the unstable pair of complex-conjugate poles are quasi-cancelled by a pair of complex conjugate zeros that lies close by, then stabilization at that location is hardly possible. A pole-zero quasi-cancellation denotes the low sensitivity to that dynamics from the analysis node or branch. We can illustrate this link between quasi-cancellations and stabilization with the aid of the simple double resonator of Fig. 4a. If the circuit is probed with a current source at node A, the unstable poles are quasi-cancelled by RHP zeros located near (Fig. 4b). Now consider the connection of a shunt resistor $R_{stab}$ at node A (in green in Fig. 4a). Decreasing the value of the shunt resistor, starting from infinite, causes the unstable poles to evolve toward the nearby zeros, so they remain in the Right-Half Plane (RHP) for any value of the shunt resistor (Fig. 4c). Consequently, the circuit cannot be stabilized with the connection of a shunt resistor at node A. On the contrary, if the circuit is probed at node B (Fig. 5a), the unstable poles are isolated on the RHP (Fig. 5b). Accordingly, if a shunt resistor $R_{stab}$ is introduced at node B and its value is swept down from infinite, the pair of unstable poles moves toward the Left-Half Plane (LHP) (Fig. 5c) and become stable at some value of the shunt resistor (500 Ω in this simple example).

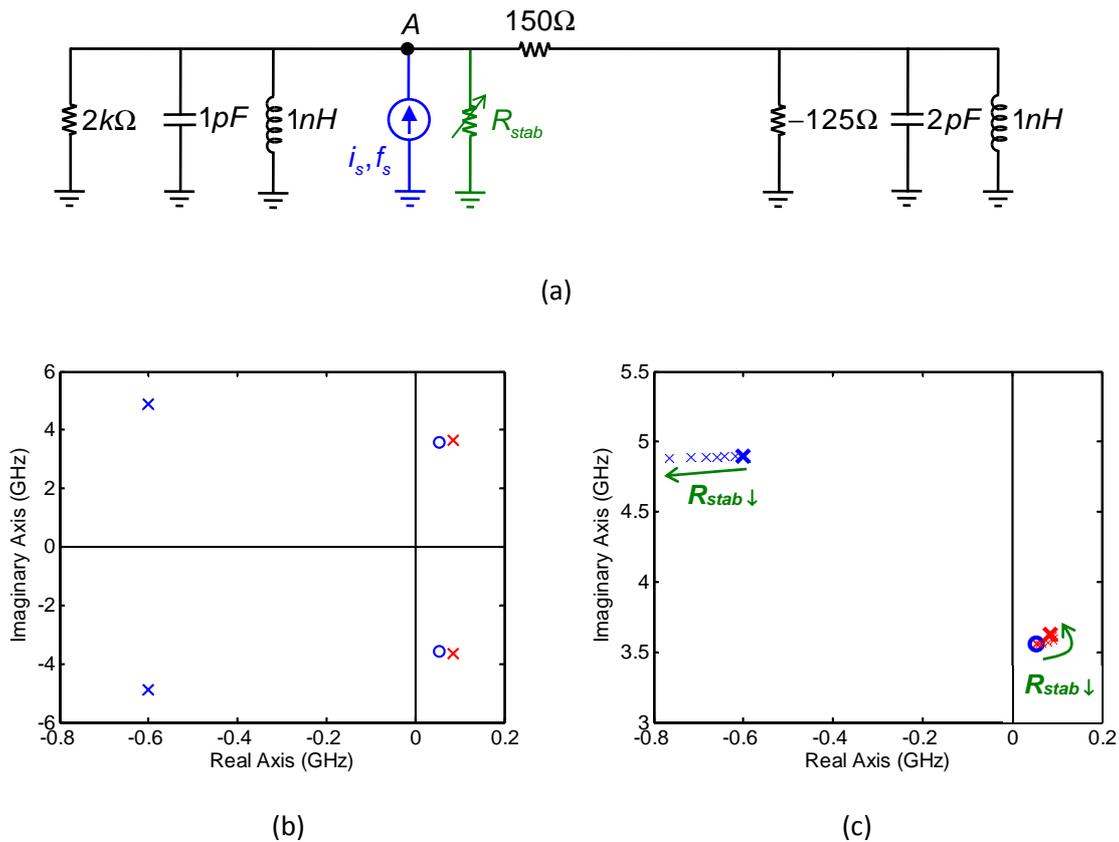

(a)

(b)                                                           (c)

Fig. 4.  a) Unstable double resonator circuit with current probe connected at node A; b) Resulting pole-zero map when the shunt resistance $R_{stab}$ is not included. The unstable poles are quasi-canceled by RHP zeros; c) Pole evolution versus $R_{stab}$. As $R_{stab}$ is swept down from infinite, unstable poles tend toward the nearby zeros and remain in the RHP for any value of the shunt resistance. For clarity, only positive frequencies of the complex plane are plotted.

These results are trivial for this simple circuit but their conclusions are general. An unstable pole-zero quasi-cancellation obtained with a current source at a given node indicates that circuit stabilization is very unlikely connecting a shunt resistor at that node. On the contrary, if unstable poles are isolated on the RHP, the amplifier can be stabilized at that analysis node through a shunt resistor. Similar reasoning applies for voltage sources connected at circuit branches and series stabilization resistors.

Developing this concept from a more accurate perspective, the work in [32] proposed pole-placement techniques from control theory (such as root-locus tracing) that can predict the exact trajectory of the critical pair of poles as the stabilization parameter varies (normally shunt or series resistances, capacitances or inductances) and thus the precise value needed to ensure stability. Those techniques apply to oscillations that build up from a dc regime.

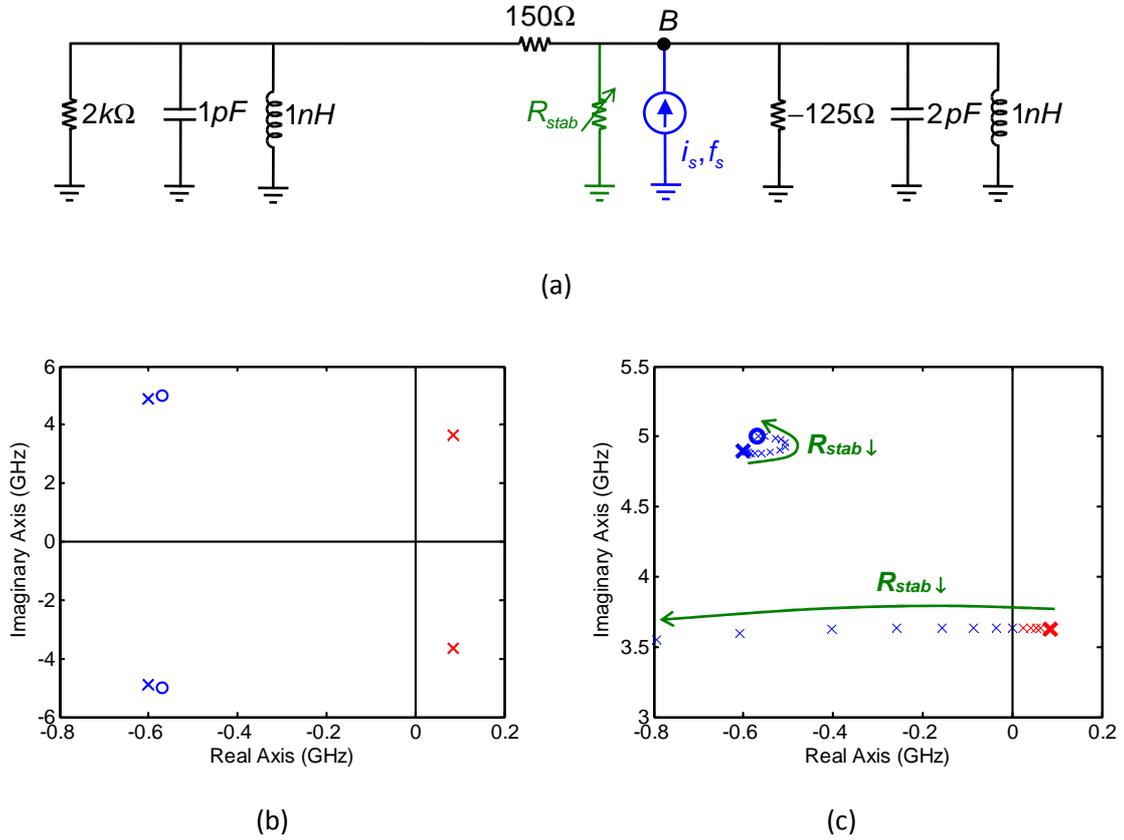

(a)

(b)                                    (c)

Fig. 5. (a) Unstable double resonator circuit with current probe connected at node B; (b) Resulting pole-zero map when the shunt resistance $R_{stab}$ is not included. The unstable poles are isolated on the RHP; (c) Pole evolution versus $R_{stab}$. As $R_{stab}$ is swept down from infinite, the pair of unstable poles moves toward the LHP and become stable as of some value of the shunt resistor. For clarity, only positive frequencies of the complex plane are plotted.

Power combining amplifiers include stabilization networks that are often too conservative. The presence of these networks causes, in the majority of cases, a degradation of the amplifier electrical performances at the operating bandwidth. The work in [33] presents a design approach to improve amplifier electrical performances through a more efficient use of pole-zero identification techniques. It allows reducing the number of stabilization networks while maintaining a sufficient stability margin. It is based on the use of a large-signal optimization process that integrates pole-zero identification from the early stages of the design. The optimization procedure is explained using a two-stage Ku-band MMIC power amplifier for telecommunication space applications that has a risk of oscillation at the divided-by-two input frequency when operated in large signal. The original prototype incorporated two kinds of stabilization networks: odd-mode parallel resistors between gates and between drains of the transistors, and parallel RC blocks connected in series to the gates of the transistors (Fig. 6). A Monte Carlo large-signal stability analysis accounting for process variability of the MMIC foundry gives the dispersion of the critical pair of poles at $f_{in}/2$ for $f_{in}$ = 12.7 GHz and at 4 dB compression (Fig. 7). Poles are stable with a reasonable stability margin that prevents oscillations due to manufacturing tolerances of the MMIC process. After application of the joint RF and stability analysis optimization, the stabilization networks can be reduced to only four inter-branch resistors as shown in Fig. 8. The same Monte Carlo analysis is performed on the optimized prototype. Dispersion of the critical poles with process variability is shown in Fig. 9. Again, poles are stable and the stability margin is satisfactory. The gain on circuit performances in the optimized prototype over the original one is given in Table 1. This approach results in an improvement of the RF performances while sufficient stability margins are preserved.

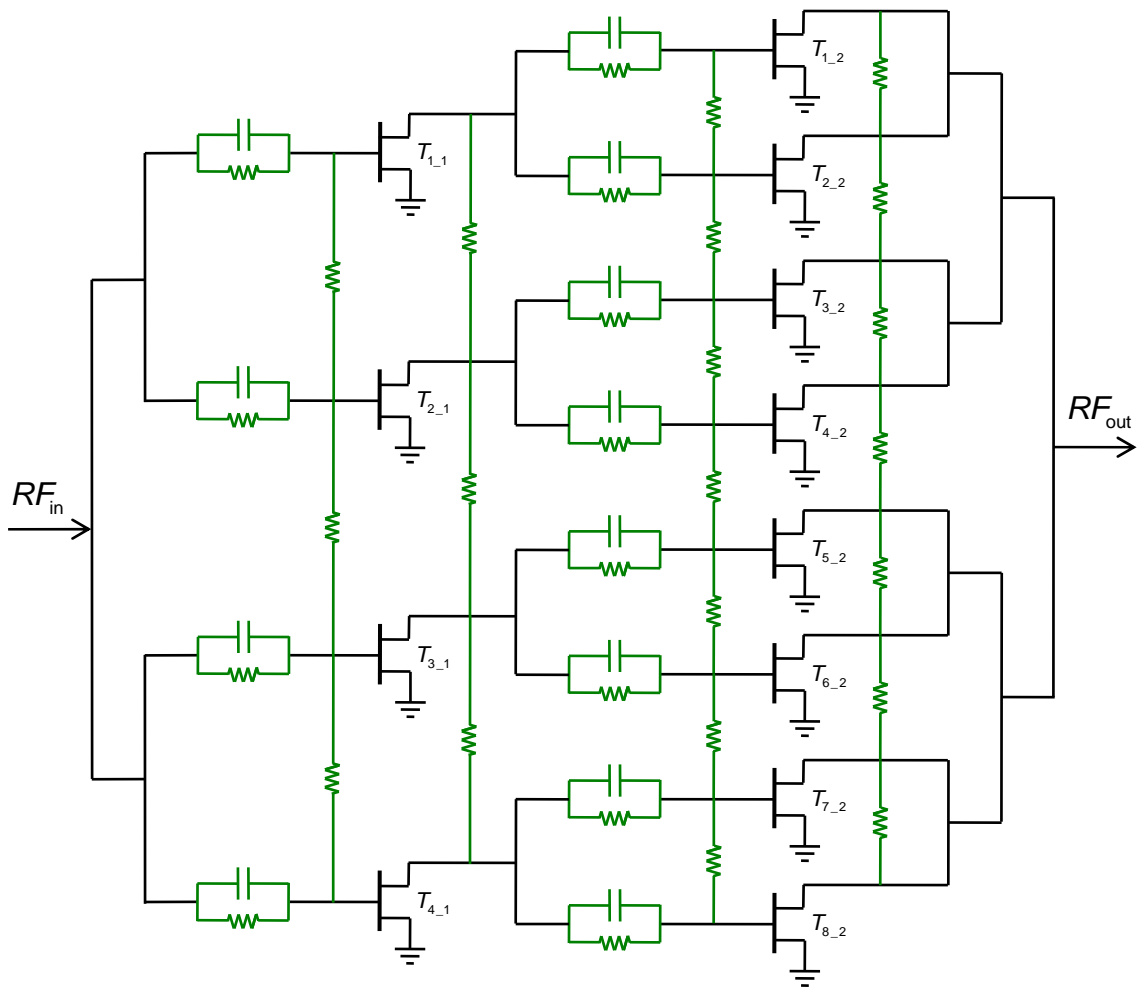

Fig. 6. Simplified electric schematics of the original design of the two-stage Ku-band MMIC power amplifier with full stabilization networks: Inter-branch resistors between gates and drains of all the transistors and RC circuits in series to the gates (from [33] with permission from Cambridge University Press).

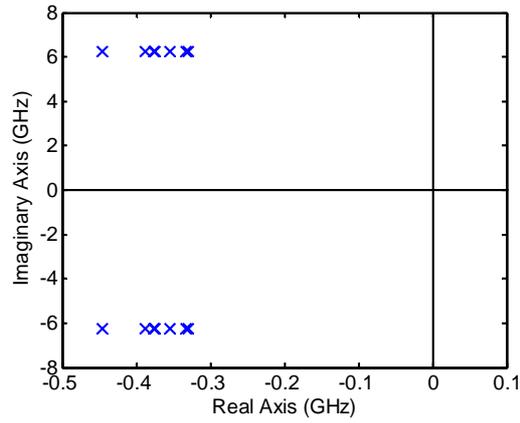

Fig. 7. Dispersion of the critical pair of complex conjugate poles at $f_{in}/2$ due to process variability obtained through Monte-Carlo analysis (100 trials) for the original circuit of Fig. 6. $f_{in}$ = 12.7 GHz and 4 dB compression (from [33] with permission from Cambridge University Press).

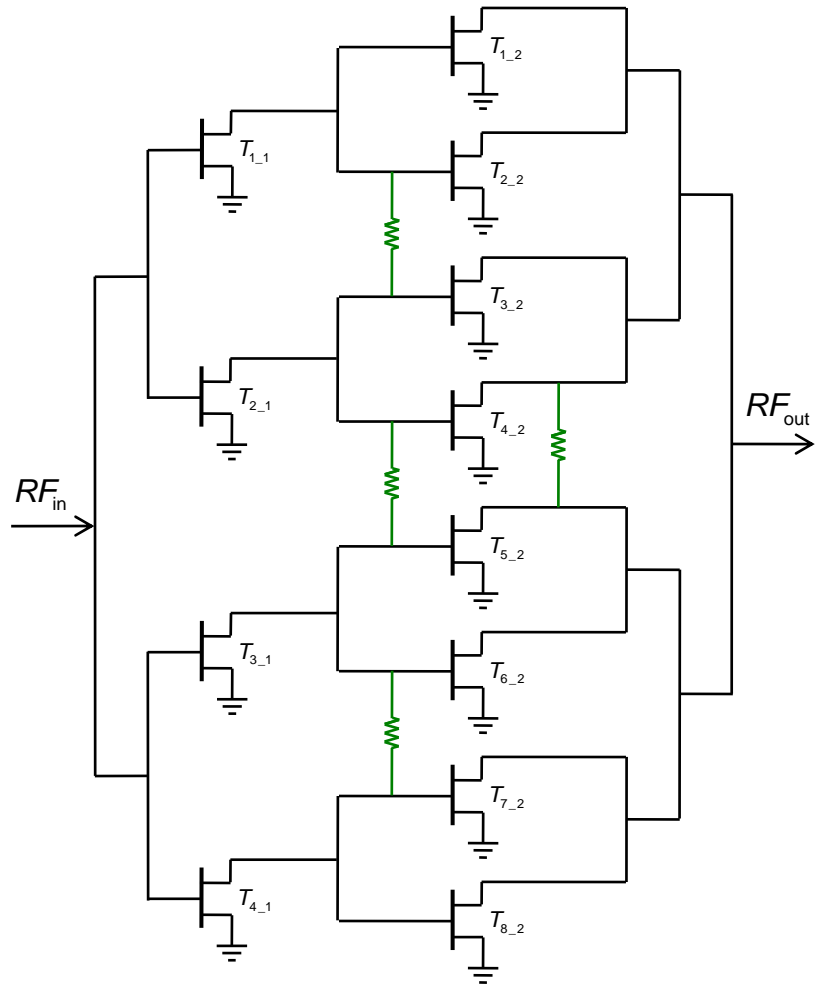

Fig. 8. Optimized electric schematic of the amplifier. Only four inter-branch resistors remain (from [33] with permission from Cambridge University Press).

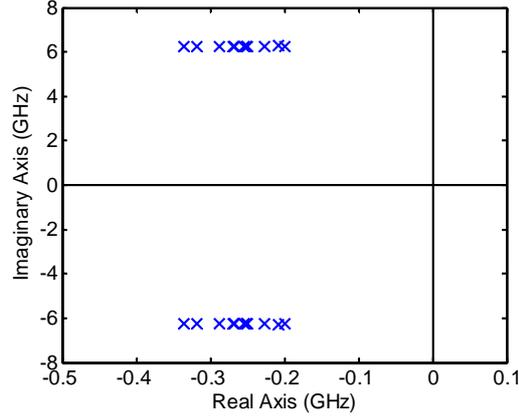

Fig. 9. Dispersion of the critical pair of complex conjugate poles at $f_{in}/2$ due to process variability obtained through Monte-Carlo analysis (100 trials) for the optimized prototype circuit of Fig. 8. $f_{in}$ = 12.7 GHz and 4 dB compression. Poles remain stable with sufficient stability margin (from [33] with permission from Cambridge University Press).

Table 1. RF performance improvements after the optimization process of the MMIC HPA [33] for different frequencies. Results are given at 2 dB compression, which is the nominal operation point for this amplifier. Nominal $P_{out}$ at 2 dB compression is 4 W.

| Frequency | ΔGain (dB) | ΔPout (mW) | ΔPAE (percentage points) |
|-----------|------------|------------|--------------------------|
| **11.2 GHz** | 3.98 | 950 | 4.6 |
| **11.7 GHz** | 2.84 | 754 | 4.07 |
| **12.2 GHz** | 2.28 | 460 | 1.9 |
| **12.7 GHz** | 2.16 | 790 | 4.55 |
| **13.2 GHz** | 3.01 | 284 | 0.72 |

*Transient optimization*

Controlling the position of unstable poles on the complex plane can be used to improve the transient characteristics for applications in which switching times are relevant. In [17], a method is proposed to improve the start-up time in oscillators for switched applications. Pole-zero identification is used to obtain the pair of complex-conjugate poles $\sigma \pm j\omega$ of the unstable solution from which the oscillation builds up. This pair of poles will dominate the initial transient of the circuit (at least while the amplitude of the growing oscillation is still small). Due to the exponential variation of the initial transient, the smaller the real part of the poles $\sigma$, the longer the transient. The technique in [17] carries out a tuning of the oscillator element so as to increase the positive real part $\sigma$ of the unstable poles while simultaneously the required oscillation frequency and first harmonic amplitude are maintained. The approach has been applied to the 2.4 GHz FET-based oscillator of Fig. 10. The source capacitance $C_S$ is the element selected to control the transient. For each value of $C_S$, the two inductances $L_G$ and $L_D$ are re-calculated to maintain the required oscillation frequency and first harmonic amplitude using an *Auxiliary Generator* technique [34]. The variation of the unstable pair of complex conjugate poles versus the set of $C_S$, $L_G(C_S)$ and $L_D(C_S)$ is shown in Fig. 11. The measured transient responses for two values of $C_S$ are plotted in Fig. 12a and are in agreement with the predictions of Fig. 11. Although the magnitude of the first harmonic is similar, the two measured waveforms have a very different harmonic content, as can be seen in the detailed plot of Fig. 12b.

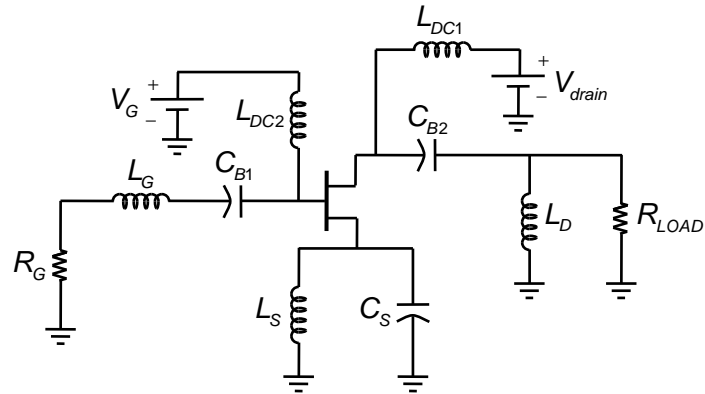

Fig. 10. Schematic of the 2.4 GHz FET-based oscillator [17]. Source capacitance $C_S$ is used to control the transient. Drain inductance $L_D$ and gate inductance $L_G$ are used to maintain the required oscillation frequency and first-harmonic amplitude for each value of $C_S$.

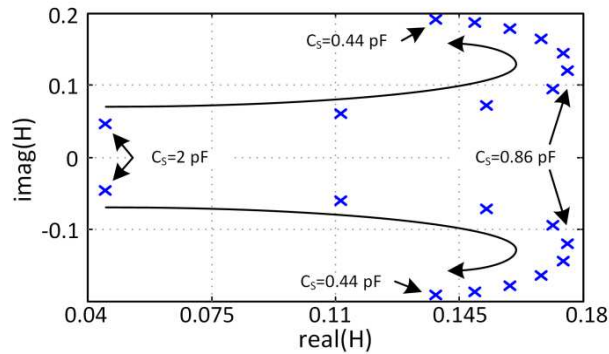

Fig. 11. Evolution of the unstable poles versus $C_S$ - $L_G(C_S)$ - $L_D(C_S)$ sets, maintaining fixed oscillation frequency and first-harmonic amplitude [17].

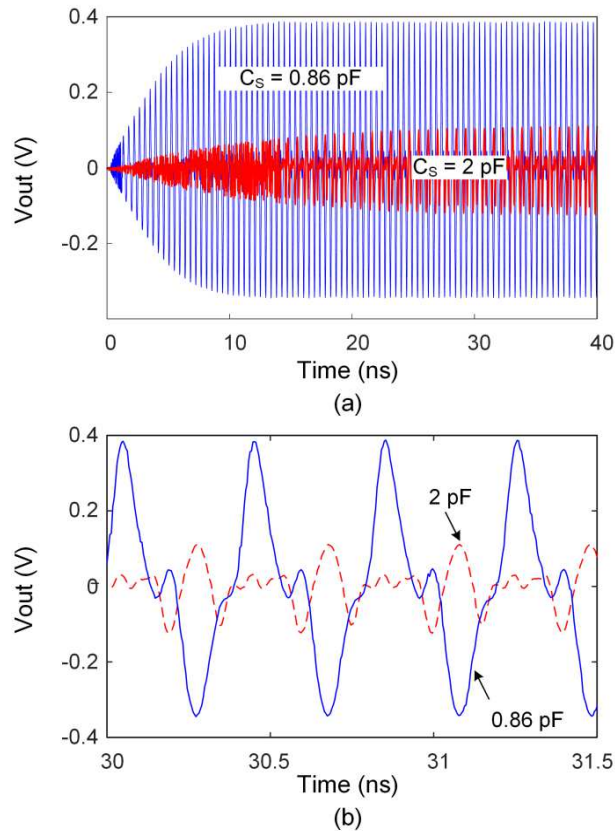

Fig. 12. (a) Measured transients for two values of the source capacitance [17]. Measurements were performed with an HP-83480 Digital Communications Analyzer. As expected from Fig. 11, the transient for $C_S$ = 2 pF is significantly longer than for $C_S$ = 0.86 pF. (b) Detailed view of the two waveforms. The harmonic content is very different, although first harmonic is similar.

*Experimental stability margin evaluation*

Although stable, Left-Half Plane (LHP) complex conjugate poles lying too close to the imaginary axis may have an undesired effect on the circuit dynamics. Having these low-damping poles may increase the risk of oscillation when circuit parameters (such as bias or frequency) or circuit external conditions (such as temperature, mounting or cabling) change, affecting the robustness of the circuit [35]. In addition, low-damping poles can be responsible for the existence of noise bumps in the output spectrum [14], [36] or long transients in switching mode amplifiers [37]. Monitoring the path of critical poles that shift dangerously toward the Right-Half Plane (RHP) when a parameter varies becomes very important to asses stability margins, at least qualitatively. This is of paramount importance when critical poles appear at low frequencies in power amplifiers, involving elements of the bias lines. Even if these critical poles do not become unstable, a location close to the imaginary axis means low damping and a high resonant effect inside the amplifier video bandwidth (Fig. 13). This resonance limits the ability of the digital pre-distortion systems to correct for intermodulation distortion in signals with large instantaneous bandwidths [38], [39].

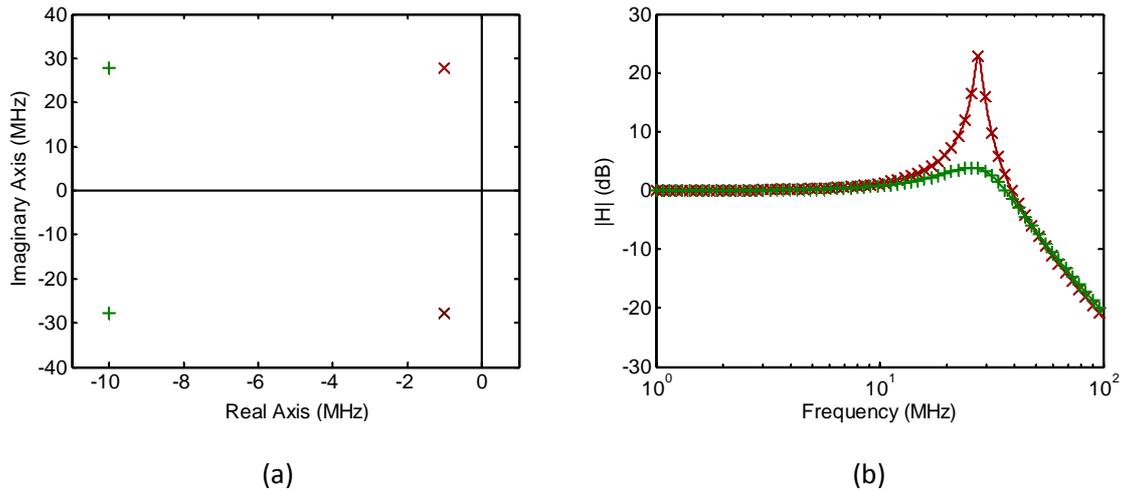

Fig. 13. Example of the effect of low-frequency dominant poles on the voltage transfer characteristics of the bias lines. (a) Two pairs of complex conjugate dominant poles are separately considered. In one case, the dominant poles (x) are notably closer to the RHP than the other (+). (b) Magnitude of voltage transfer function H associated to each pair of poles. A high resonant effect is obtained for the poles located close to the RHP.

An experimental method to characterize the low-frequency stability margins of microwave amplifiers has been proposed in [35]. The experimental approach is especially useful for those situations in which simulation is not fully reliable either because accurate non-linear models of the active devices are not available or because there is an incomplete electrical description of the circuit. The low-frequency dominant poles are extracted from reflection coefficient measurements performed at observation ports that are specifically included in the gate and drain bias paths of the circuit to get access to the low-frequency dynamics. This method has been used in [40] to optimize the design of the bias lines in terms of video bandwidth, relative stability margins and voltage transfer characteristics in a GaN power amplifier (Fig. 14). The low-frequency dominant poles have been characterized versus different circuit parameters. In Fig. 15.a we show how a pair of complex conjugate poles scatters when modifying three parameters: gate and drain bias voltages and input power. The result is a cloud of poles, all of them stable. However, there are some parameter configurations for which the pair of poles lies dangerously close to the RHP. A redesign of the bias networks (in this case an increase of the resistance in series with the low-frequency decoupling capacitors in gate and drain) manages to shift the critical pair of poles leftwards, increasing stability margin and reducing their resonant effect on the amplifier video bandwidth (Fig. 15.b).

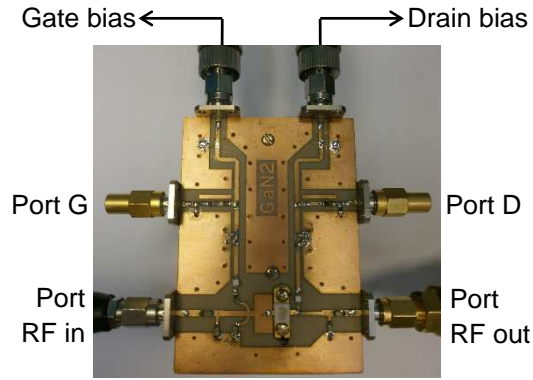

Fig. 14.  Wideband [0.4-1.7 GHz] power amplifier based on GaN HEMT device CGH40010F from CREE fabricated in microstrip hybrid technology [40]. Two observation ports G and D are added in the bias paths, in series with the RC access networks.

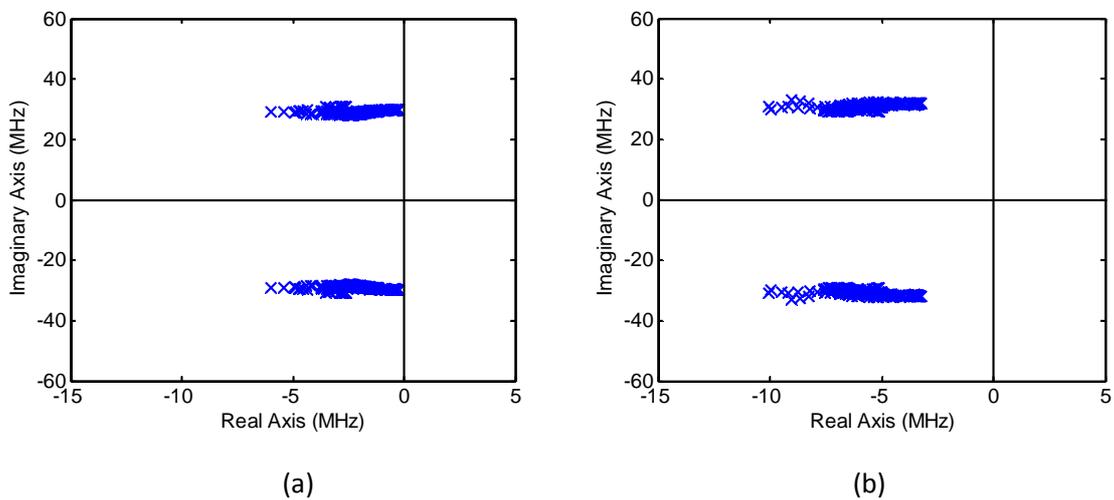

(a)                                                (b)

Fig. 15.  Clouds of stable low-frequency critical poles obtained by varying three circuit parameters (gate bias, drain bias and input power) for two different values of the resistance in series with the low frequency decoupling capacitors in gate and drain [40]. (a) Series resistance of 20 Ω. For some parameter configurations poles are dangerously close to the RHP. (b) Series resistance of 50 Ω. The cloud of poles is shifted leftwards, increasing stability margin and reducing the resonant effect of the poles on the amplifier video bandwidth, as according to Fig. 12.

*Bifurcation analysis*

Global stability and bifurcation analysis of non-linear microwave circuits is another context in which pole-zero identification is usually utilized as a complementary technique. Bifurcations are qualitative changes in the circuit solution when a parameter is varied continuously [34]. The tracing of bifurcation loci serves to delimit the operation bands of circuits of autonomous nature such as VCOs, injection locked oscillators and frequency dividers. A comprehensive description of most relevant local bifurcations in microwave circuits and their link to the crossing of the poles through the imaginary axis on the complex plane is given in [41]. Bifurcation-detection techniques (such as those based on the *Auxiliary Generator* technique [34]) through harmonic-balance simulations are often combined with the tracing of pole trajectories on the complex plane to study the stability of solution curves [42]-[47], which provides an insightful knowledge of the dynamic behavior of the autonomous circuit. As an example, in [43], a closed-form formulation for the optimized design of coupled oscillators has been presented. Pole-zero identification is used to show the reduction of the stability margin when the number of coupled oscillators increases (Fig. 16). As the system gets larger it becomes more sensitive to discrepancies between oscillator elements and is more likely to become unlocked.

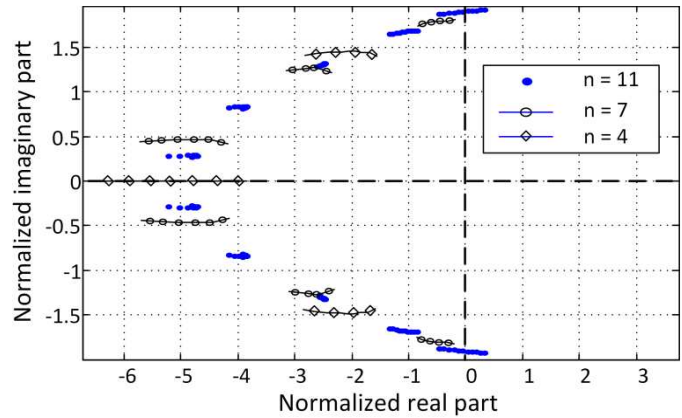

Fig. 16. Pole locus showing the reduction of the stability margin, due to discrepancies between oscillator elements, when the number $n$ of oscillator elements increases in a coupled oscillator system [43]. Three cases have been considered $n = 4$, 7 and 11. As the system gets larger, the poles approach the imaginary axis and the system eventually unlocks for $n = 11$.

In [45], an in-depth stability and bifurcation analysis of self-oscillating quasi-periodic solutions is presented and applied to the self-oscillating power amplifier (SOPA) of Fig. 17. Through a theoretical study, these authors demonstrate that pole-zero identification can be applied to the stability analysis of quasi-periodic states. As an example, stable and unstable SOPA solution curves are represented in Fig. 18 in terms of the oscillation amplitude for different bias conditions. Stability analyses of the quasi-periodic solutions, marked as A and A' in Fig. 18, are shown in Fig. 19, where the real part of the dominant poles versus input power is plotted.

Bifurcation analysis combined to pole-zero identification can be found in many other papers that carry out detailed investigations of the dynamic behavior of non-linear autonomous circuits, such as multi-resonant free-running oscillators [42]; varactor-based frequency dividers and multipliers [44], ring oscillators [46], and non-linear transmission line based oscillators [47].

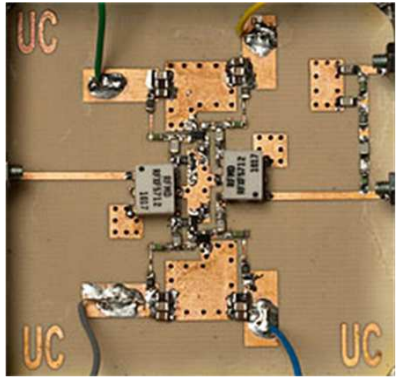

Fig. 17.   Photograph of the SOPA based on an RF Class-D power amplifier topology [45]. The transistor used is an ATF33143 HFET (Avago Technologies). Switching frequency is in the order of 750 MHz.

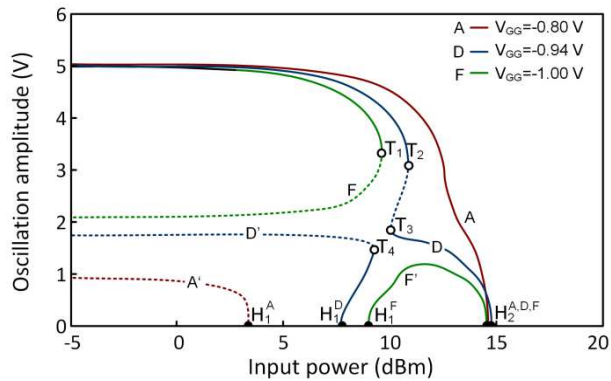

Fig. 18.   SOPA solution curves, represented in terms of the oscillation amplitude versus input power for different gate bias voltages [45]. Stable sections are traced in solid line and unstable sections are traced in dashed line. $T_1$ to $T_4$ are turning points. Hopf bifurcation points are labeled with H.

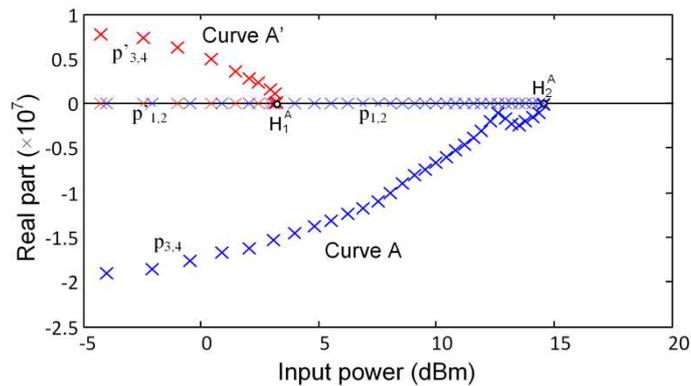

Fig. 19.   Stability analysis of the quasi-periodic solutions A and A' from Fig. 18 [45]. Evolution of the real part of the dominant poles versus input power for gate bias $V_{GG}$ = -0.8V. Note how the stability of A is correctly predicted by the pole evolution even at a $P_{in}$ of about 12.5 dBm where the curve has an almost vertical slope.

*Verification of the Rollet proviso*

When investigating the conditional/unconditional stability of a two-port network versus source and load terminations, pole-zero identification is also a perfect complement to Rollet stability criterion [48]. The aim of the well-known Rollet stability criterion is to determine if the linear two-port exhibits negative resistance at its ports for some values of passive source and load terminations, which could lead to circuit oscillation. Satisfying the Rollet criterion will only imply unconditional stability if the linear two-port does not contain any internal unstable loops not observable from the external ports. This condition is formulated in a proviso that has to be fulfilled for the Rollet stability criterion to be sufficient. Verifying the proviso is important in circuits with complex topologies, such as multistage amplifiers, because of the lack of observability of internal dynamics from the input and output ports. The proviso can be verified by confirming that there are no intrinsic unstable poles when ports are loaded with open and short terminations. This condition can be easily checked applying pole-zero identification at internal circuit nodes with the circuit loaded with open and short terminations.

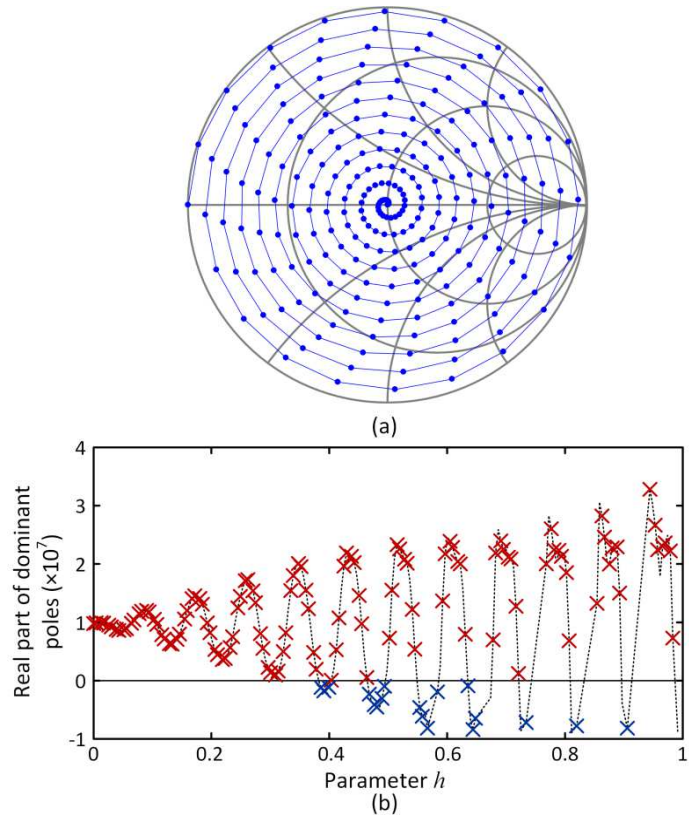

Fig. 20.    (a) Continued coverage of the Smith Chart following a spiral curve: $\Gamma(h) = 0.999he^{j(2N\pi+\pi)h}$ with $h$ being the single sweeping parameter ($0 \leq h \leq 1$) and $N$ = 11. (b) Checking out the proviso on the power amplifier prototype in [49] for all the points of the $h$-sweep show internal unstable poles under short-circuit terminations at the relevant sidebands.

Things are more complicated if we want to generalize Rollet stability criterion to the analysis of the unconditional stability versus load terminations in large-signal regimes. From a general perspective, this is an incommensurable multidimensional problem because, for each input power $P_{in}$, negative resistance at frequency $f_s$ will depend on load terminations at $Kf_{in}$ and at side bands $Kf_{in} \pm f_s$, with $K$ being the number of relevant harmonics and $f_{in}$ the fundamental frequency of the input signal. A rigorous generalization of the Rollet stability criterion to large-signal periodic regimes under output mismatch effects is given in [49]-[50]. It is applied to practical cases in which the output filter of the amplifier allows the consideration of only two (or three at most) relevant sidebands at the load termination port. The fundamental upper and lower sidebands ($f_{in} \pm f_s$) are normally considered as the virtual observation ports for the Rollet analysis. In this case, verification of the Rollet proviso is more demanding because it must be checked for each passive termination at $f_{in}$ with the two sideband frequencies in all possible combinations of short-circuit and open-circuit terminations. In other words, for each possible combination of short-circuit and open-circuit terminations at sideband frequencies, pole-zero identification at internal nodes has to be carried out versus variations of load termination at $f_{in}$. These variations must cover the whole Smith Chart and could be implemented with a double sweep in magnitude and phase. However, in doing so, we get disconnected circles as we explore the Smith Chart, which impedes taking advantage of the continuity properties of the harmonic balance simulation. A smart solution is given in [49]. The analysis follows a single spiral curve on the Smith Chart that depends on a single parameter (Fig. 20). The proviso is checked out for all the points of the spiral curve.

**Pole-zero Identification with Vector Fitting for stability analysis of microwave amplifiers**

As explained in the previous section, pole-zero identification carried out at several nodes/branches of the circuit provides valuable information on where and how to improve circuit stability. When this is done using a commercial tool, as in [32], the multiple frequency responses are identified as a sequence of independent Single-Input, Single-Output (SISO) pole-zero analyses. In theory, all the individual SISO transfer functions of a linear (or linearized) system share a common denominator that is the characteristic equation. This means that, except for exact pole-zero cancellations, we should get the same poles (only zeros will change) at all the analysis ports when performing multiple independent SISO analyses. However, in practice, some disagreement might appear for high damping poles or for poles that are quasi-canceled by zeros. Consequently, it seems more convenient to use Multiple-Input Multiple-Output (MIMO) frequency identification algorithms that maintain a common denominator for frequency responses obtained at different observation ports.

In this direction, an approach based on Vector Fitting [51], [52] that performs the frequency domain identification of a vector of transfer functions using a common denominator has been recently proposed in [53] to analyze the stability of multistage amplifiers. An advantageous aspect of using a Vector Fitting based algorithm for stability analysis of microwave circuits is that the identified transfer functions $H_n(s)$ are described using a partial fraction representation ($H_n$ is the $n^{th}$ component of a vector of transfer functions):

$$H_n(s) = \sum_{k=1}^{N} \frac{r_{n,k}}{s - p_k} + D \qquad (1)$$

where $p_k$ are the system poles, common to all the $H_n$, $r_{n,k}$ is the residue corresponding to pole $p_k$ and transfer function $H_n$, and $D$ is the direct gain. On the contrary, existing commercial pole-zero identification tools for microwave amplifier stability analysis [28], are based on least-square optimization and the transfer functions are represented as the ratio of two polynomials:

$$H_n(s) = \frac{a_0 + a_1 s + a_2 s^2 + \cdots + a_N s^N}{b_0 + b_1 s + b_2 s^2 + \cdots + b_N s^N} \qquad (2)$$

The expansion in partial fractions (1) has a numerical advantage when identifying large bandwidths that involve very high frequencies and high transfer function orders $N$, compared to the use of the ratio of two polynomials (2). Actually, using (2) results in excessive large numbers due to the terms in powers of $s$, leading to an ill-conditioned numerical problem. It is often necessary to divide the original problem into sub-bands when a ratio of two polynomials is used for the identification [23]. However, with a partial fraction representation as in (1), large numbers are avoided and the frequency response identification is numerically better conditioned [51]. This has an important consequence: large bandwidth responses can be identified without dividing the frequency response into narrower sub-bands and algorithms for automatic order selection of the transfer function are simplified and can be more effective.

*Residue Analysis*

Fixing the unstable behavior of a circuit by adding or modifying a stabilization parameter (such as a stabilization resistance) at a particular node or branch depends on the sensitivity to the unstable dynamics from that spot. This stabilization capability could be qualitatively estimated from the observation of the pole-zero quasi-cancellations on the complex plane, as explained in the previous section. Isolated pairs of complex-conjugate poles reveal an observation port with high sensitivity, while pole-zero quasi-cancellations indicate low sensitivity to that dynamics from the observation port.

Taking advantage of the representation in partial fractions (1), a procedure based on residue analysis is developed in [53] to quantify the "sensitivity" of the identification obtained at a particular circuit location. In order to do that, a normalized factor $\rho_{n,k}$ is defined to quantify the relative effect of a pair of resonant complex conjugate poles $p_k$, $p_k^*$ on the transfer function $H_n$:

$$\rho_{n,k} = \frac{\left| H_n(j\omega_r) \right|}{\left| H_n(j\omega_r) - H_{n,k}(j\omega_r) \right|} \qquad (3)$$

where $\omega_r$ is the resonant frequency of the poles $p_k$, $p_k^*$ and $H_{n,k}$ represents their contribution to the transfer function $H_n$:

$$H_{n,k}(s) = \frac{r_{n,k}}{s - p_k} + \frac{r_{n,k}^*}{s - p_k^*} = \frac{\left( r_{n,k} + r_{n,k}^* \right) s - \left( r_{n,k} p_k^* + r_{n,k}^* p_k \right)}{\left( s - p_k \right)\left( s - p_k^* \right)} \qquad (4)$$

Poles that are identified at highly sensitive locations will show high values of $\rho$. On the contrary, poles identified at low sensitivity locations are quasi-cancelled by zeros and will have small $\rho$ values. A similar approach to detect spurious poles with low effect on the frequency response can be found in [54].

The advantages of analyzing $\rho$ for a set of poles obtained at different nodes or branches are twofold:

On the one hand, we can use the computation of $\rho$ to minimize the adverse effects of over-modeling in the pole-zero fitting process [23]. If the pole-zero identification is carried out using a very strict optimization goal, we could end up by fitting the numerical noise existing in the simulated frequency response. This can lead to the onset of pole-zero quasi-cancellations that are non-physical as shown in Fig. 21a. These non-physical poles will have extremely low values of $\rho$, no matter the circuit location at which we obtain the transfer function. On the contrary, poles representing the actual circuit dynamics will have significantly larger values of $\rho$ for at least some analysis ports (nodes or branches). Therefore, we are able to discriminate numerical pole-zero quasi-cancellations by analyzing their values of $\rho$ corresponding to transfer functions obtained at different nodes/branches. This approach has been applied to the example of Fig. 21a in order to clean up the pole-zero map from numerical pole-zero quasi-cancellations, resulting in the map of Fig. 21b where only physical poles remain.

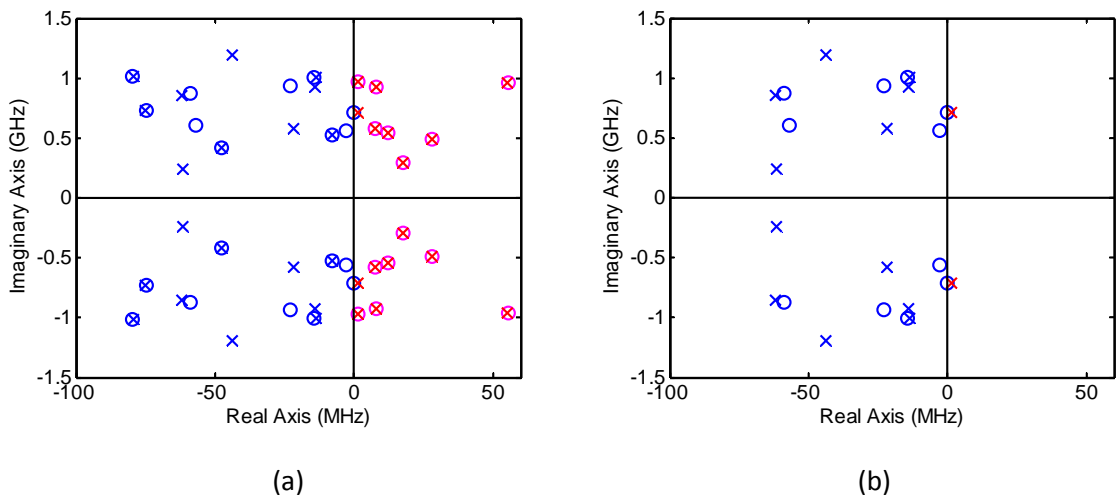

(a)                              (b)

Fig. 21.  (a) Example of over-modeling due to a too strict optimization criterion (the criterion in this case is the phase difference between the data and the model and has been set to 0.01 degrees). Non-physical pole-zero quasi-cancellations are modeling the numerical noise present in the frequency response, complicating the correct interpretation of the results. (b) Quasi-cancellations due to over-modeling have been eliminated through an analysis of the $\rho$ values of all the identified poles. The system is unstable and a proper detection of unstable poles is reached.

On the other hand, by performing the analysis at different nodes and branches and analyzing $\rho$ for the identified poles, we can classify the different locations of the circuit with respect to their influence in the unstable dynamics and thus deduce the most sensitive places to stabilize the circuit.

A simple example of this approach is given in [53], where the three-stage amplifier based on GaAs FET transistors (FLU17XM) of Fig. 22 is considered. Experimentally, this prototype showed an oscillation at 3.5 GHz for the nominal bias point (Fig. 23). A MIMO stability analysis is performed for that bias condition. Eight observation nodes, corresponding to gate and drain terminals of the four transistors (labeled $n_1$ to $n_8$ in Fig. 22), are taken into account. Since the amplifier has a power combining structure in its third stage, excitation of odd and even modes has been considered. To excite the even mode, two in-phase small-signal current sources are simultaneously applied to nodes $n_5$ and $n_6$ (same for nodes $n_7$ and $n_8$). To excite the odd-mode, two 180° out-of-phase small-signal current sources are simultaneously applied to nodes $n_5$ and $n_6$ (same for nodes $n_7$ and $n_8$). Eventually, a total of eight frequency responses are identified with a common denominator. The resulting poles plotted on the complex plane are shown in Fig. 24. An unstable pair of complex conjugate poles is obtained at 3.3 GHz. Next, computation of factor $\rho$ of this unstable pair is performed for the eight frequency responses. Results are graphically shown in Fig. 25. We can clearly infer that the oscillation is taking place in the third stage with an odd-mode nature because its corresponding $\rho$ values are several orders of magnitude larger than at the rest of the analyzed nodes and modes. To eliminate this oscillation, inter-branch resistors between gates or drains of third stage transistors can be introduced (Fig. 22). Note that the value of $\rho$ for the odd-mode frequency response between the gates of the third stage is larger than the value obtained for the odd-mode frequency response between the drains. This suggests that the optimum place for a stabilization resistor is between the gates of the transistors of the third-stage. That was experimentally confirmed. An inter-branch resistor of 5.1 KΩ between the gates was enough to eliminate the observed instability at 3.5 GHz, while a much lower value (910 Ω) was needed between the drains for stabilization. Note that low inter-branch resistances can impact amplifier performances when there are appreciable symmetry imbalances due to technological dispersion.

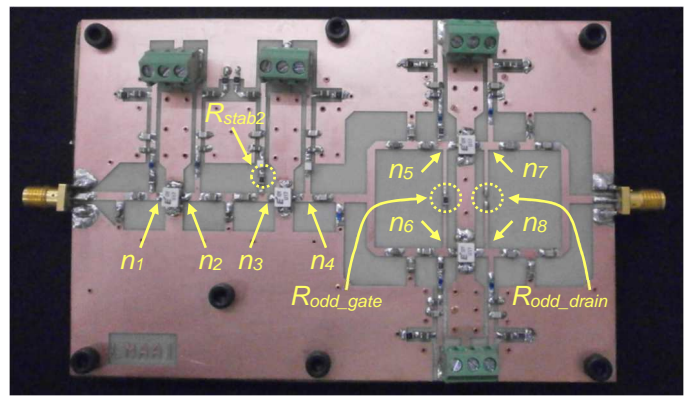

Fig. 22.    Photograph of the hybrid 3-stage GaAs FET-based power amplifier prototype [53].

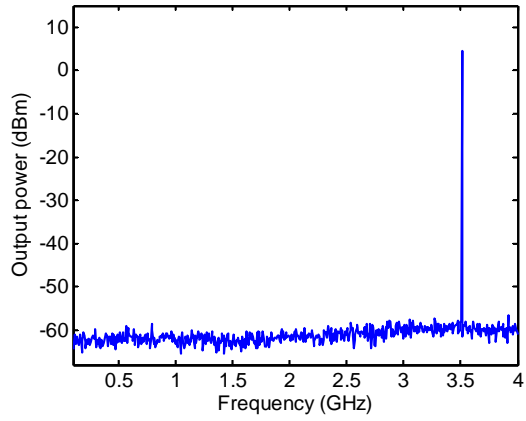

Fig. 23.  Measured output power spectrum of the three-stage PA showing an oscillation at 3.5 GHz when biased at nominal conditions [53].

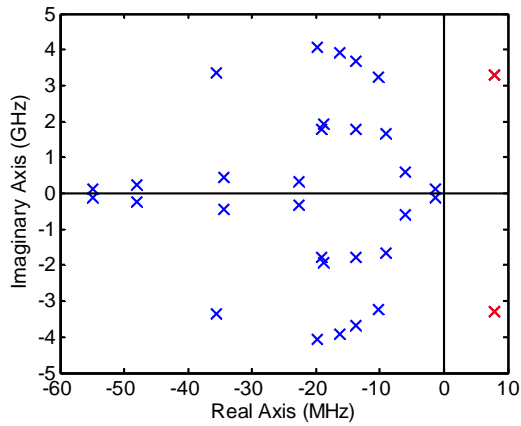

Fig. 24.  The MIMO pole-zero identification of the 8 significant frequency responses results in the detection of a pair of unstable complex-conjugate poles at 3.3GHz [53].

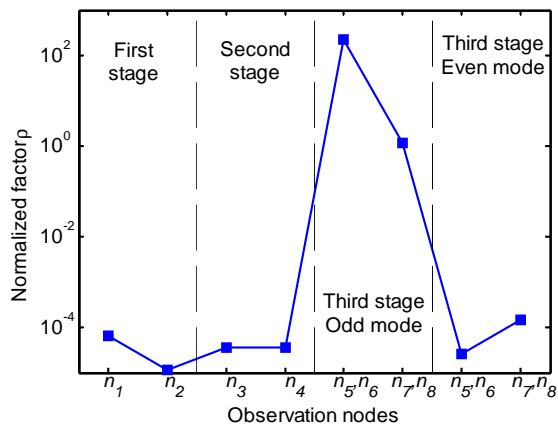

Fig. 25.  Residue analysis of unstable poles at 3.3GHz indicating that the instability is an odd-mode oscillation at the third stage [53].

**Conclusions and Final Recommendations**

Pole-zero identification is a valuable tool for gaining insight into the dynamics of active microwave circuits. It is simple, intuitive and easy to use. Pole-zero identification is particularly robust in detecting resonant poles in the analyzed frequency band. This is very convenient since critical resonances are associated with pairs of complex conjugate poles with low damping. However, in order to obtain reliable results, pole-zero identification has to be properly used within its validity limits. We next give some final recommendations for a use of pole-zero identification tools that leads to consistent interpretation of stability results.

- Do not rely on simple pass/fail results. Whenever you find a pair of complex conjugate poles with positive part, trace the pole trajectories versus relevant circuit parameters (such as bias, or input power) until you see the bifurcation point (the crossing from the RHP to the LHP). This provides valuable information for stabilization and discards numerical errors.

- Once critical poles are detected, use parametric analyses to monitor the pole trajectory on the LHP as well. Sometimes models are not accurate enough and stable poles close to the RHP warn you about an oscillation risk.

- Do not use very strict optimization goals in the identification. Otherwise you will end up by fitting the numerical noise in the frequency response (over-modeling) giving rise to unphysical pole-zero quasi-cancellations. It is advisable to perform the analysis at a few nodes/branches (one per stage in the case of multistage amplifiers and close to a transistor preferably) in a MIMO configuration with a relaxed tolerance for the fitting. This will detect the critical resonances avoiding the over-modeling problems. Besides, it determines where the origin of the instability is, which is of paramount importance when trying to fix an undesired oscillation. Once the best location is determined we can focus on that node/branch and go on with further SISO parametric analyses. A MIMO analysis does not mean every single node/branch at the input and output of all active devices present in the circuit. Since pole-zero identification is a very powerful optimization tool, selecting a few nodes or branches is enough. Typically, analyzing at one node or branch per stage in the case of multi-stage amplifiers is adequate. This is important to save simulation time when computing the frequency responses in the commercial simulator, especially relevant when analyzing large-signal periodic steady states.

- Finally, always keep in mind that in order to get reliable results the key point is the calculation of the frequency response in the electrical simulator, not so the identification that follows. A frequency response with large numerical errors, with truncation noise, or simply incorrect due to inaccurate electrical models, may lead to wrong conclusions about the stability of the system.

**Acknowledgements**

Part of the work referred here has been supported by the French Space Agency (CNES) (projects R-S10/TG-0001-019 and R-S14/TG-0001-019); by a joint funding of a Ph.D. research grant from CNES and Thales Alenia Space – France; and by project TEC2015-67217-R (MINECO/FEDER).

**Closed-loop frequency response**

The first step of the stability analysis is the obtaining of a closed-loop frequency response by introducing a current probe in parallel at a circuit node or a voltage probe in series at a circuit branch. These figures show the equivalence between the current or voltage perturbation and the general scheme of a feedback system.

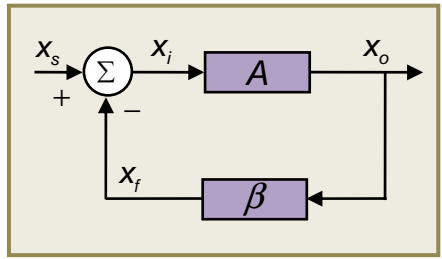

a) General feedback system.

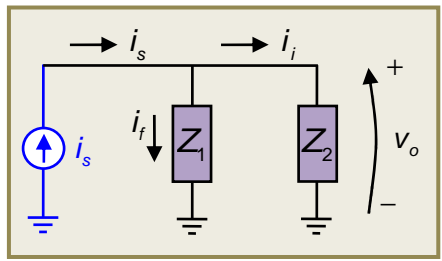

b) Basic shunt/shunt feedback topology, where the input signal is the current $i_s$ injected at a particular node and the output signal is the voltage $v_o$.

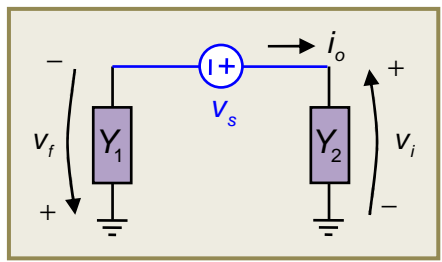

c) Basic series/series feedback topology, where the input signal is the voltage $v_s$ inserted at a particular circuit branch and the output signal is the current flowing through the branch $i_o$.

**Side bar 2**

**Stability in large-signal operation**

An amplifier, originally stable under DC and small signal conditions, can exhibit spurious oscillations from a certain value of the input power drive. Those oscillations can have two origins. One is the combined effect of a feedback with a gain expansion versus power. This typically appears in transistors biased in class B and deep AB, where gain expansion versus input power is common. The other possibility is the negative resistance exhibited by a nonlinear capacitance (as those included in the models of the active devices) pumped by the input drive. This negative resistance plus a resonance effect facilitates the parametric generation of sub-harmonic frequencies, as in a pumped varactor diode. A comprehensive explanation of the physical origin of this parametric effect can be found in [S1]. A typical example of this parametric oscillation is the frequency division-by-two that is so common in power amplifiers with several transistors combined in parallel.

[S1] – A. Grebennikov and N.O. Sokal, *Switchmode RF Power Amplifiers*, Elsevier, Oxford, UK, 2007

**Even-mode / odd-mode oscillations**

Due to symmetry, amplifiers with power combining structures can have different oscillation modes. In the simplest case of two transistors in parallel (Fig. S1), two different oscillation modes are possible: an even mode in which the two transistors oscillate in-phase and an odd mode in which the two transistors oscillate 180º out of phase. The kind of oscillation mode can be determined using pole-zero analysis. The even and odd modes can be individually excited and observed injecting two current probes with appropriate phases at gates of the parallel transistors: both probes in-phase excite primarily the even mode while two probes 180º out of phase excite mainly the odd-mode (Fig S1). Note that the odd-mode oscillation is not observable at the power division/combination nodes, which are virtual grounds for the odd mode. Injecting the current probe at those nodes would turn out in exact pole-zero cancellations.

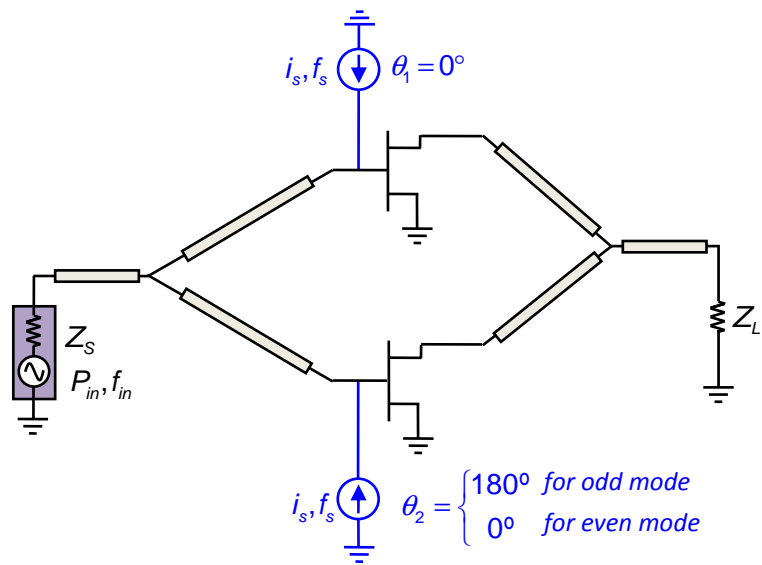

Fig. S1. Individual excitation of oscillation modes: Odd-mode ($\theta_1 = 0$º, $\theta_2 = 180$º) and even-mode ($\theta_1 = \theta_2 = 0$º).